\documentclass{iau} 
\usepackage{graphicx}

\title[IAUS332~~The chemical connection between 67P/C-G and IRAS~16293-2422] 
{The chemical connection between\\67P/C-G and IRAS~16293-2422}

\author[M. N. Drozdovskaya, E. F. van Dishoeck, M. Rubin, J. K. J{\o}rgensen \& K. Altwegg]   
{Maria Nikolayevna Drozdovskaya$^{1}$,
 Ewine F. van Dishoeck$^{2,3}$,
 Martin Rubin$^{4}$,
 Jes Kristian J{\o}rgensen$^{5}$
 \and Kathrin Altwegg$^{4,1,6}$}

\affiliation{$^1$Center for Space and Habitability (CSH), Universit{\"a}t Bern \\ Sidlerstrasse 5, CH-3012 Bern, Switzerland \\ email: {\tt maria.drozdovskaya@csh.unibe.ch} \\[\affilskip]
$^2$Leiden Observatory, Leiden University \\ PO Box 9513, NL-2300 RA Leiden, The Netherlands \\ email: {\tt ewine@strw.leidenuniv.nl} \\[\affilskip]
$^3$Max-Planck Institut f{\"u}r Extraterrestrische Physik (MPE) \\ Giessenbachstrasse 1, 85748 Garching, Germany \\[\affilskip]
$^4$Physikalisches Institut, Universit{\"a}t Bern \\ Sidlerstrasse 5, CH-3012 Bern, Switzerland \\ email: {\tt martin.rubin@space.unibe.ch} \\[\affilskip]
$^5$Centre for Star and Planet Formation \\ Niels Bohr Institute \& Natural History Museum of Denmark, University of Copenhagen \\ {\O}ster Voldgade 5–7, DK-1350 Copenhagen K., Denmark \\ email: {\tt jeskj@nbi.ku.dk} \\[\affilskip]
$^6$email: {\tt kathrin.altwegg@space.unibe.ch}}

\pubyear{2017}
\volume{332}  
\jname{Astrochemistry VII – Through the Cosmos from Galaxies to Planets}
\editors{Maria Cunningham, Tom Millar \& Yuri Aikawa, eds.}
\begin{document}

\maketitle

\begin{abstract}
The chemical evolution of a star- and planet-forming system begins in the prestellar phase and proceeds across the subsequent evolutionary phases. The chemical trail from cores to protoplanetary disks to planetary embryos can be studied by comparing distant young protostars and comets in our Solar System. One particularly chemically rich system that is thought to be analogous to our own is the low-mass IRAS~16293-2422. ALMA-PILS observations have made the study of chemistry on the disk scales ($<100$~AU) of this system possible. Under the assumption that comets are pristine tracers of the outer parts of the innate protosolar disk, it is possible to compare the composition of our infant Solar System to that of IRAS~16293-2422. The \textit{Rosetta} mission has yielded a wealth of unique in situ measurements on comet 67P/C-G, making it the best probe to date. Herein, the initial comparisons in terms of the chemical composition and isotopic ratios are summarized. Much work is still to be carried out in the future as the analysis of both of these data sets is still ongoing.
\end{abstract}

\firstsection 
\section{The Assembly Line}
Star and planet formation are two inseparable, intertwined processes that initiate upon the gravitational collapse of a prestellar core. The infalling gas and dust of the core are primarily partitioned between the young protostar and the protoplanetary disk. The leftover core materials remain in the form of an envelope around the young star-disk system during the embedded phase of star formation (e.g., \cite{Shu1977}). Some gas and dust are also expelled from the system via the action of outflows.

The deepest layer of a protoplanetary disk, the midplane, is the site of most-efficient build up of larger refractory bodies. It is still unclear exactly when grain growth from micron to millimeter and centimeter sizes occurs; however, meter- and kilometer-sized bodies will undoubtedly only be able to form in the midplanes, where the densities are high-enough for such masses to accumulate. This makes midplanes the birth places of comets and planetary embryos (e.g., \cite{WilliamsCieza2011}). In turn, it is collapsing core materials that feed the disk midplanes. What remains to be understood is the degree of processing or pristinity of the refractory and volatile components.

One potential means of tackling this question is by considering the chemical connection embedded in volatiles across the pre- and proto-stellar phases of star formation. It is possible to study the gaseous and solid volatiles in prestellar cores by means of observations in the submillimeter (e.g., ALMA) and infrared (e.g., \textit{Spitzer}) frequency ranges. However, in the sub-surface layers of protoplanetary disks the densities are high and temperatures are low. This implies that most volatiles are in the solid phase and obscured from view by the thick dust disk. This makes direct midplane observations almost impossible (except less abundant isotopologes of highly volatile species such $^{13}$C$^{18}$O; \cite{Zhang2017}) and imposes the need for indirect tracers, such as comets.

Comets are thought to form alongside planetary embryos in the disk midplane. It is possible that they remain relatively small, of km-size, due to being formed in the outer disk, where the densities are lower and thus there being less material available for growth of refractories. It is thought that Kuiper Belt comets remain in the distant parts of the disk until their orbit is perturbed by a giant planet's gravitational field, which can only occur in a mature planetary system, such as our own modern day Solar System. Oort Cloud comets have been suggested to have formed closer-in, e.g., near Jupiter, and then slung out to large distances during the early formation phases of our Solar System (e.g., \cite{AHearn2011, MummaCharnley2011}).

As cometary data are only available for our Solar System (the field of exo-comets still remains a future perspective) and direct observations of the pre- and proto-stellar phases are unavoidably only towards other forming stars, some assumptions need to be made. First of all, it is assumed that other low-mass protostellar systems are analogous to our innate protosolar nebula (PSN). Secondly, comets are seen as pristine tracers of the PSN, i.e., excluding any processing that may have occurred during their time in a mature planetary configuration.

\section{\textit{Rosetta} and 67P/C-G}
Comets have primarily been studied via remote sensing and ground-based observations, and also dedicated \textit{Stardust} and \textit{Deep Impact} missions (as reviewed in \cite{AHearn2011}). This picture changed in 2014 when the \textit{Rosetta} mission approached, went into orbit and carried out a landing operation on the comet 67P/Churyumov-Gerasimenko (67P/C-G, hereafter). Between August 6th, 2014 and September 30th, 2016 the \textit{Rosetta} orbiter remained in orbit around the comet and surveyed its surface and coma with an array of instruments. The lander \textit{Philae} was in operation for about 3 days upon its triple landing on November 12th, 2014. The \textit{Rosetta} mission has rendered the most unique and diligent dataset on cometary science yet with its in situ measurements.

Jupiter-family comets (JFCs) are short-period comets with low inclinations and aphelia near Jupiter's orbit, and are thought to originate from the Kuiper Belt. 67P/C-G is a JFC of around $\sim 4 \times 3 \times 2$~km in size with a period of $6.44$~yr, an aphelion of $5.68$~AU and a perihelion of $1.24$~AU with the latest occurring on August 13th, 2015. Due to the inclination of its orbit relative to the orbital plane of the Solar System, it experiences seasonal changes. The Northern hemisphere experiences a long warm summer of $\sim 5.5$~yr, while the Southern hemisphere undergoes an intense short summer between its inbound equinox on May 5th, 2015 and its outbound equinox on April 9th, 2016. The \textit{Rosetta} orbiter monitored its target during all three of these major events.

\section{ALMA-PILS and IRAS~16293-2422~B}
Low-mass protostars have been studied extensively with single-dish facilities, interferometers and space missions. The latest powerful successor is the Atacama Large Millimeter/submillimeter Array (ALMA). ALMA has been used to carry out the Protostellar Interferometric Line Survey (PILS; project-id: 2013.1.00278.S, PI: Jes K.
J{\o}rgensen) towards the low-mass binary system IRAS~16293-2422 in Band~$7$ ($329-363$~GHz). The survey is unique not only due to its wide frequency coverage, but also its high angular resolution of $\sim60$~AU, spectral resolution of $\sim0.2$ km s$^{-1}$ and sensitivity of $4-5$~mJy~beam$^{-1}$~km~s$^{-1}$. Of particular interest is source B, which has narrow lines with full width at half maximum (FWHM) of $\sim1$~km~s$^{-1}$ and its face-on disk orientation. This reduces line blending and facilitates line identification, thus revealing the chemistry of its disk-region (\cite{Jorgensen2016}).

\section{The chemical connection: composition}
Jointly, the \textit{Rosetta} data on 67P/C-G and the ALMA observations of IRAS~16293-2422~B, are some of the best data sets on a PSN midplane tracer and a PSN analogue. Hence, the two can be used to explore the chemical connection spanning the earliest stages of star formation and the protoplanetary building blocks, much better than was done in the past (e.g., \cite{Bockelee-Morvan2000}). One comparative aspect is the chemical composition of the two targets.

The dominant volatiles of 67P/C-G have been shown to form two groups, those following H$_{2}$O and those following CO$_{2}$. The first group includes O$_{2}$ and NH$_{3}$, while the second group contains CO, H$_{2}$S, CH$_{4}$ and HCN. The partitioning is based on the density profiles pre- and post-perihelion (from January to August 2016), and the rate of decrease post-perihelion (\cite{Gasc2017}). Water is the dominant volatile and the peak production rate was seen three weeks after perihelion (\cite{Hansen2016}). All of these species have been detected towards IRAS~16293-2422 and are also well-known constituents of interstellar ice based on infrared observations against protostellar and background sources (\cite{Boogert2015}).

The detection of O$_{2}$ at a level of $\sim4\%$ relative to H$_{2}$O remains a surprising result (\cite{Bieler2015}), as it is a species rarely detected under interstellar conditions. The search for this molecule towards IRAS~16293-2422 is still ongoing with dedicated ALMA Cycle 4 data under analysis at the moment (Taquet et al., 2018, in prep.). Astrochemical models are suggesting that such a high quantity of O$_{2}$ can only be formed through grain surface chemistry under prestellar core conditions, albeit at a lightly elevated $\sim20$~K temperature than normal. Chemistry within the protoplanetary disk is not efficient enough for the production of such a high abundance (\cite{Taquet2016}). Other mechanisms such as radiolysis may also contribute to the prestellar O$_{2}$, although this mechanism overproduces HO$_{2}$, H$_{2}$O$_{2}$ and O$_{3}$ (\cite{Mousis2016}).

A classical biomarker species, CH$_{3}$Cl, has been detected in the coma of 67P/C-G with the ROSINA instrument. The molecule has also been observed towards IRAS~16293-2422~A and B. The two detections, unfortunately, imply that CH$_{3}$Cl cannot be considered a biomarker. However, the Cl-bearing molecule remains the dominant organohalogen in the Earth's atmosphere and is seen at similar proportions towards 67P/C-G and IRAS~16293-2422~B, $\sim0.007-6 \times 10^{-4}$ and $\sim7 \times 10^{-5}$ relative to CH$_{3}$OH, respectively (\cite{Fayolle2017}).

Comet 67P/C-G has also been shown to carry glycine, atomic phosphorus and an entire zoo of complex organic molecules. It has been seen that it is abundant in S-bearing molecules, but poor in N-bearing molecules (\cite{LeRoy2015, Altwegg2016, Calmonte2016, Altwegg2017b}, Wampfler et al., 2018, in prep.). IRAS~16293-2422 is also rich in a variety of O-, N- and S-bearing complex organic species (e.g., \cite{Jorgensen2016}); however, its complexity is more comparable to that of a petting farm rather than a zoo. It cannot be entirely excluded that the lack of detections of more complex molecules is not due to sensitivity issues as a result of complicated partition functions, which lead to weaker lines, and the larger distances between the observer and the target.

ROSINA measurements have estimated that $\sim57\%$ of the volatile sulfur is incorporated into H$_{2}$S, $\sim27\%$ into atomic S and then smaller portions into SO$_{2}$, SO, OCS, H$_{2}$CS and other molecules. Sulfur-chain species, including S$_{2}$, S$_{3}$, S$_{4}$, have been detected on 67P/C-G and are likely fragments of the semi-refractory S$_{8}$. S-bearing organics, CH$_{3}$SH and C$_{2}$H$_{6}$S have also been measured in the coma (\cite{Calmonte2016}). All these cometary molecules containing a single S atom have been also detected towards IRAS~16293-2422, except for atomic S and C$_{2}$H$_{6}$S. Thus, in terms of diversity, there is an agreement. However, the latest analysis are showing that IRAS~16293-2422~B is much more abundant in OCS; and that H$_{2}$S is by-far not as much of a dominant S-carrier as is seen in the comet (Drozdovskaya et al., 2017, under rev.).

\section{The chemical connection: isotopic ratios}
Another clue of the chemical trail across evolutionary stages lies in the isotopic ratios. The D/H ratio on 67P/C-G has now been measured in singly and doubly deuterated water, and in H$_{2}$S. The values are $\left( 1.05\pm0.14 \right) \times 10^{-3}$ for HDO/H$_{2}$O, $\left( 1.8\pm0.9 \right) \times10^{-2}$ for D$_{2}$O/HDO and $\left( 1.2\pm0.3 \right) \times10^{-3}$ for HDS/H$_{2}$S, respectively (\cite{Altwegg2017a}). This implies that the ratio is molecule-dependent and may have been set at different evolutionary times corresponding to the time of formation of each molecule. The D/H as measured in terms of the singly deuterated water is higher than that of other JFCs, but closer to the range of protostellar envelope values. The best estimate in terms of singly deuterated water towards IRAS~16293-2422~B based on interferometric ALMA and SMA observations is $\left( 9.2\pm2.6 \right) \times 10^{-4}$ for HDO/H$_{2}$O(\cite{Persson2013}). It is important to compare with observations at high spatial resolution, as water is present on large and small scales in such star-forming regions and the D/H ratio varies therein. Of relevance for the cometary comparisons, are the small disk-scales, which can only be reached with interferometers, where directly thermally desorbed ices are seen.

The higher D/H measured in the doubly deuterated water has been explained in astrochemical models via the process of deuteration occurring at a later stage of formation in the upper layers of icy grain mantles (\cite{Furuya2016, Furuya2017}). This remains to be tested from the observational perspective towards IRAS~16293-2422. The D/H as measured in terms of H$_{2}$S and complex organics, for example, remains the topic of future work as the analysis of the ROSINA and PILS data is still ongoing. Further work is also required to compare isotopic ratios in volatiles of other species, including, C, O, N, S and Cl.

\section{So far, what does this tell us?}
With some of the best cometary and protostellar data at hand, the \textit{Rosetta} measurements of 67P/C-G and ALMA-PILS observations of IRAS~16293-2422~B, it seems that the PSN, at least in part, contains prestellar materials. Qualitative comparisons in terms of the chemical composition of both targets are indicating that other low-mass systems, such as IRAS~16293-2422, may be analogous to our young Solar System. Quantitative comparisons and the detailed studies of isotopic ratios remain the subject of future research (Drozdovskaya et al., 2018, in prep.).


\end{document}